\documentclass{aastex}

\RequirePackage{color}

\newcommand{\emaila}{diw.arun@gmail.com}

\begin{document}

\title{The orientation of elliptical galaxies}
\shorttitle{The orientation of elliptical galaxies}
\shortauthors{D. K. Chakraborty and A. K. Diwakar}

\author{D. K. Chakraborty\altaffilmark{1} and A. K. Diwakar\altaffilmark{1,2}}

\altaffiltext{1}{School of Studies in Physics and Astrophysics, Pt. Ravishankar Shukla University, Raipur, Chhattisgarh - 492010, India.}
\altaffiltext{2}{\emaila}

\begin{abstract}
We determine the orientations of the light distribution of individual elliptical galaxies by combining the profiles of photometric data from the literature with triaxial models. The orientation is given by a Bayesian probability distribution. The likelihood of obtaining the data from a model is a function of the parameters describing the intrinsic shape and the orientation. Integrating the likelihood over the shape parameters, we obtain the estimates of the orientation. We find that the position angle difference between the two suitably chosen points from the profiles of the photometric data plays a key role in constraining the orientation of the galaxy. We apply the methodology to a sample of ten galaxies. The alignment of the intrinsic principle axes of the NGC 3379, 4486 and NGC 5638 are studied.
\end{abstract}


\section{Introduction}
Intrinsic shapes and orientations of the individual elliptical galaxies have been investigated by Binney (1985), Tenjes et al. (1993), Statler (1994a,b), Bak and Statler (2000), Statler (2001) and Statler et al. (2004). These authors have used the kinematical data and the photometric data, and have used the triaxial models with the density distribution $\rho(m^{2})$, where $m^{2} = x^{2} + \frac {y^{2}} {p^{2}} + \frac {z^{2}} {q^{2}}$ with constant axial ratios $p$ and $q$. It was shown analytically that the projected density of such a distribution $\rho(m^{2})$ is stratified on similar and co-aligned ellipses (Stark 1977, Binney 1985). Statler (1994a,b, hereafter S94a, S94b) uses (apart from the kinematial data) a constant value of ellipticity, which is an average over a suitably chosen range of the radial distance, for the shape and the orientation estimates. The shape estimates are robust, and are described by a pair of shape parameters, namely the short to long axial ratio of the light distribution and the triaxiality of the mass distribution. The orientations are described by the Euler angles which define the orientation of the galaxies relative to the line of sight.

The methodology is based on Bayesian statistics, and is described in S94a, S94b, and more recently in Statler et al. (2004). The likelihood of obtaining the data from a model in a chosen orientation is a function of the orientation parameters and the shape parameters. The likelihood is multiplied by a model for the parent distribution of these parameters (the prior), and integrated over the parameters that one is not trying to constrain. The normalized result is the posterior probability distribution for the parameters of interest. Probability of the shape is obtained by integrating over the uninteresting orientation parameters, and likewise, probability of the orientation is obtained by integrating over the shape parameters. The shape and orientation thus evaluated do not constitute a simultaneous fit to the data ; they are, in fact, mutually exclusive (S94a).

Chakraborty et al. (2008, hereafter C08) have demonstrated that the variation in the intrinsic shapes of the light distribution of elliptical galaxies can be determined by using the photometric data alone. They use the models with varying axial ratios (Chakraborty 2004) and the triaxial models discussed by deZeeuw and Carollo (1996). These models exhibit ellipticity variation and position angle twist. They use ellipticities $\epsilon_{in}$, $\epsilon_{out}$ and the position angle difference $\Theta_{out} - \Theta_{in}$ at two suitably chosen points $R_{in}$ and $R_{out}$ from the profiles of the photometric data of the galaxies. The intrinsic shape is described by the parameters $(q_{0}, q_{\infty}, T_{0}, T_{\infty})$, where $ q_{0}, q_{\infty} (T_{0}, T_{\infty}) $ are short to long axial ratios (triaxialities) at small and at large radii, respectively. They find that the best constrained shapes parameters are the axial ratios $q_{0}$ and $q_{\infty}$, and the absolute value of the triaxial difference $T_{d}$ defined as $|T_{d}| = |T_{\infty}-T_{0}|$. The methodology adopted in C08 is essentially the same as described in S94a with necessary modifications to suit the requirements.

We now take up the complementary problem. We determine a statistical estimate of the orientation of the light distribution of the galaxies using photometric data. We consider the usual polar coordinates $(\theta^{'},\phi^{'})$ of the line of sight, with respect to the principle axes of the galaxy, to define the orientation. We use ellipticities $\epsilon_{in}$, $\epsilon_{out}$ and the position angle difference  $\Theta_{d} = \Theta_{out} - \Theta_{in}$, as in C08, and calculate the likelihood of obtaining the data from a model in a chosen orientation. 

We consider a flat parent distribution $(q_{0}, q_{\infty}, T_{0}, T_{\infty}, \theta^{'}, \phi^{'})$ = constant (flat prior) and multiply it with the likelihood to obtain the posterior density. Integrating the posterior density over the shape parameters, we obtain the marginal posterior density $\cal{P}(\theta^{'}, \phi^{'})$ describing the statistical estimate of the orientation. It is necessary that the marginal posterior density should be likelihood dominated, and not prior dominated. The prior, which is considered as flat in the present investigation, may be replaced by an improved or modified prior when the results of a large sample of galaxies are available. Modified prior is estimated and used in Bak and Statler (2000).

Although it is the posterior density which constitutes the result of the Bayesian estimates, some statistical summary is also useful for its description. Area of the parameter space enclosing $68\%$ of the total probability, which can be interpreted as $1\sigma$ error bar is one such useful quantity. A working definition of a good likelihood dominated estimate is that the $2\sigma$ region (area enclosing $95\%$ of the total probability) should be less than $50\%$ of the total parameter space (S94a). Another useful quantity is the location $(\theta^{'}_{p}, \phi^{'}_{p})$ of the maximum probability. 

In the present investigation, we consider the triaxial galaxies under study to have very simple photometric structure. Each one of them is considered as a triaxial distribution of one component luminous matter. Thus, we ignore complications, such as dust, shells or disks etc. In this context, we mention that dust lanes or disks are sometimes used to constrain the orientations and shapes (see, eg. Tenjes et al. 1993, S94a). We also assume that the axes of each triaxial galaxy are intrinsically aligned, and therefore, the orientation as determined by using the observed data at a small $R_{in}$ and at a large $R_{out}$ gives the orientation of the entire galaxy. We shall examine this assumption in section 5.

Determination of the intrinsic shape and the orientation using photometry is important because the number of galaxies with good photometry is many more than those with good kinematics. We find that photometry can constrain the orientation of the light distribution of the triaxial galaxies. We find that the position angle difference plays a key role in constraining the orientation. The most probable line of sight of a galaxy with very small $|\Theta_{d}|$ lies close to one of the principle planes, although the estimates are not very much likelihood dominated. The orientation estimates are likelihood dominated and are well constrained for galaxies with large $|\Theta_{d}|$, typically around $4^{o}.0$ or more. The marginal posterior density is symmetrical in 4 pairs of octants in all cases of small and large $|\Theta_{d}|$ (These points are described in detail in sections 3 and 4).

Section 2 presents the triaxial mass models. Section 3 presents the orientations of NGC 7619 and two synthetic galaxies, which establish our methodology. Section 4 discusses the orientations of a sample of galaxies. In section 5, we examine the alignment of the intrinsic principle axes of NGC 3379, 4486 and 5638. Results and a discussion are presented in section 6.
\section{Model}
We use models which are described in C08. The models are triaxial generalizations of the spherical $\gamma$-models $(0 \leq \gamma < 3)$ of Dehnen (1993). These models have cusp at the centre and the cumulative mass converges. The projected surface density corresponding to $\gamma = 1.5$ most closely resembles to the de Vaucouleurs $R^{\frac {1} {4}}$ law, and we concentrate to $\gamma = 1.5$ models only. However, it is realized that the shape and orientation estimates are not very sensitive to the radial profile of the projected density (S94b).

A triaxial generalization of Dehnen's spherical density distribution $\rho(r)$ is made triaxial by considering the distribution $\rho(r)$ and replacing $r$ by $M$, where $M^{2} = x^{2} + \frac {y^{2}} {P^{2}} + \frac {z^{2}} {Q^{2}}$. Here, $(x,y,z)$ are the usual Cartesian coordinates, $r$ is the spherical radial coordinate and 
\begin{eqnarray}
P^{-2}(M) = \frac{\beta b^{2}p_{0}^{-2}+M^{2}p^{-2}_{\infty}}{\beta b^{2}+M^{2}}, \\
Q^{-2}(M) = \frac{\beta b^{2}q_{0}^{-2}+M^{2}q^{-2}_{\infty}}{\beta b^{2}+M^{2}}.
\end{eqnarray}

The constant density surfaces are coaxial and coaligned ellipsoids with varying axial ratios $(P,Q)$, which are $(p_{0},q_{0})$ at small radii and are $(p_{\infty},q_{\infty})$ at large radii. In above, $b$ is the scale length and $\beta > 0$ is a parameter. The parameter $\beta$ alters the values of $(P,Q)$ in the intermediate region. These models are presented in Chakraborty (2004) and are modified in C08. We shall refer to these as $M^{2}$ models.

Another form of triaxial generalization of Dehnen's model is presented in deZeeuw and Carollo (1996), and is modified in C08. Rewriting $\rho(r)$ as $f(r)$, the triaxial model is density distribution of the form
\begin{eqnarray}
\rho=f(r)-[g(r)+g_{1}(r)]Y^{0}_{2}+[h(r)+h_{1}(r)]Y^{2}_{2},
\end{eqnarray}
where $\rho$ is the density in the usual spherical polar coordinates $(r,\theta,\phi)$, $g(r)$ and $h(r)$ are two suitably chosen radial functions, $Y_{2}^{0} = 3/2\cos^{2}{\theta}-1/2$ and $Y_{2}^{2} = 3\sin^2{\theta}\cos{2 \phi}$ are the usual spherical harmonics and 
\begin{eqnarray}
g_{1}=\alpha \frac{Lr_{1}}{4\pi (r+r_{2})^{7}}[30rr_{2}],\\
h_{1}=\alpha \frac{Lr_{3}}{4\pi (r+r_{4})^{7}}[30rr_{4}].
\end{eqnarray}
In above, $g(r)$, $h(r)$ and the parameters $r_{1}..r_{4}$ are the same as in deZeeuw and Carollo. $g_{1}(r)$ and $h_{1}(r)$ are introduced in C08, which for choices of parameter $\alpha > 0$, alter the profiles of $\epsilon$ and $\Theta$ in the intermediate region. $L$ is the luminosity of the model. We shall refer to these as $fgh$ models.

Both the $M^{2}$ and the $fgh$ models form a family of four parameters models, and are fixed by a choice of axial ratios $(p_{0}, q_{0}, p_{\infty}, q_{\infty})$. The triaxialility $T_{0}$ and $T_{\infty}$ are related to the axial ratios at small and at large radii by
\begin{eqnarray}
T_{0} = \frac{1-p_{0}^{\gamma}}{1-q_{0}^{\gamma}} \ , \ T_{\infty} = \frac{1-p_{\infty}^{4}}{1-q_{\infty}^{4}}.
\end{eqnarray}
for the $fgh$ models, and by 
\begin{eqnarray}
T_{0} = \frac{1-p_{0}^{2}}{1-q_{0}^{2}} \ , \ T_{\infty} = \frac{1-p_{\infty}^{2}}{1-q_{\infty}^{2}}.
\end{eqnarray}
for the $M^{2}$ models. To fix up the scale length $b$ of the triaxial models, we consider the value of effective radius $R_{e} = 1.28b$ of the spherical models. The effective radius of the triaxial models depends on the axial ratios, as well as on the viewing angles. However, such changes are small for $\gamma$ models (deZeeuw and Carollo, 1996) and are neglected. For a choice of $(q_{0}, T_{0}, q_{\infty}, T_{\infty})$, we take various values of $\beta$ and $\alpha$ to calculate the probability. The unweighted sum of the probability using these ensemble of models gives a model independent estimate of the shape and orientation. 
\begin{table*}
\centering
\caption{Variation of projected properties of the models with orientation.}
\begin{tabular}{ccccccc}
  \hline
& \multicolumn{3}{c}{ $M^{2}$ Model with $(q_{0}, T_{0}, q_{\infty}, T_{\infty})$} & \multicolumn{3}{c}{ $fgh$ Model with $(q_{0}, T_{0}, q_{\infty}, T_{\infty})$} \\
& \multicolumn{3}{c}{$= (0.9, 0.1, 0.5, 0.2)$ \& $\theta'=30^{o}$ } & \multicolumn{3}{c}{$= (0.7, 0.3, 0.9, 0.1)$ \& $\theta'= 60^{o}$} \\\hline
$\phi'$ & $\epsilon_{in}$ & $\epsilon_{out}$ & $\Theta_{d}$ & $ \epsilon_{in}$ & $\epsilon_{out}$ & $\Theta_{d}$ \\ \hline
10 & 0.04&	0.04&	13.54& 0.01   & 0.01  & -1.11  \\
20 & 0.04&	0.06&	14.40& 0.01    &0.01  & -1.53 \\
30 & 0.04&	0.08&	11.07& 0.01   & 0.02  & -1.33 \\
40 & 0.05&	0.10&	7.92&  0.02   & 0.02  & -0.96 \\
50 & 0.05&	0.12&	7.20&  0.02   & 0.03  & -0.64 \\
60 & 0.06&	0.13&	3.64&  0.03   & 0.03  & -0.41 \\
70 & 0.06&	0.14&	3.60&  0.03   & 0.03  & -0.24\\
80 & 0.06&	0.14&	1.02&  0.03   & 0.04  & -0.11 \\
90 & 0.06&	0.14&	0.01&  0.03   & 0.04  & 0.00\\
100& 0.06&	0.14&	-0.37&	0.03  &  0.04 & 0.11\\
110& 0.06&	0.14&	-2.96&	0.03  &  0.03 & 0.24\\
120& 0.06&	0.13&	-4.60&	0.03  &  0.03 & 0.41\\
130& 0.05&	0.12&	-6.49&	0.02  &  0.03 & 0.64\\
140& 0.05&	0.10&	-7.85&	0.02  &  0.02 & 0.96\\
150& 0.04&	0.08&	-10.80&	0.01  &  0.02 & 1.33\\
160& 0.04&	0.06&	-13.10&	0.01  &  0.01 & 1.53\\
170& 0.03&	0.04&	-12.86&	0.01  &  0.01 & 1.11\\ \hline
\end{tabular}
\end{table*}

We study the variation of the projected properties of the models with orientation. We consider (1) a $M^{2}$ model with $\beta = 1$ and $(q_{0}, T_{0}, q_{\infty}, T_{\infty}) = (0.9, 0.1, 0.5, 0.2)$ and (2) a $fgh$ model with $\alpha = 0$ and $(q_{0}, T_{0}, q_{\infty}, T_{\infty}) = (0.7, 0.3, 0.9, 0.1)$ and calculate the 'photometric' data at $R_{in}=0.25R_{e}$ and $R_{out}=1.40R_{e}$ $(R_{e} = 1.28)$. To project these models we take a fixed value of $\theta^{'}$ and a varying $\phi^{'}$. The variations in $\epsilon_{in}, \epsilon_{out}$ and $\Theta_{d}$ as a function of $\phi^{'}$ are presented in Table 1. The variation in the position angles difference $\Theta_{d}$ is noteworthy. The variation is large, and it changes sign across $\phi^{'}=90^{o}, 180^{o}, 270^{o}$ and $360^{o}$. We find that this has a significant role in constraining the orientation. On the other hand, changes in ellipticities with $\phi'$ is relatively small.
\begin{figure}
\centering
\includegraphics[width=12.0cm, angle=0]{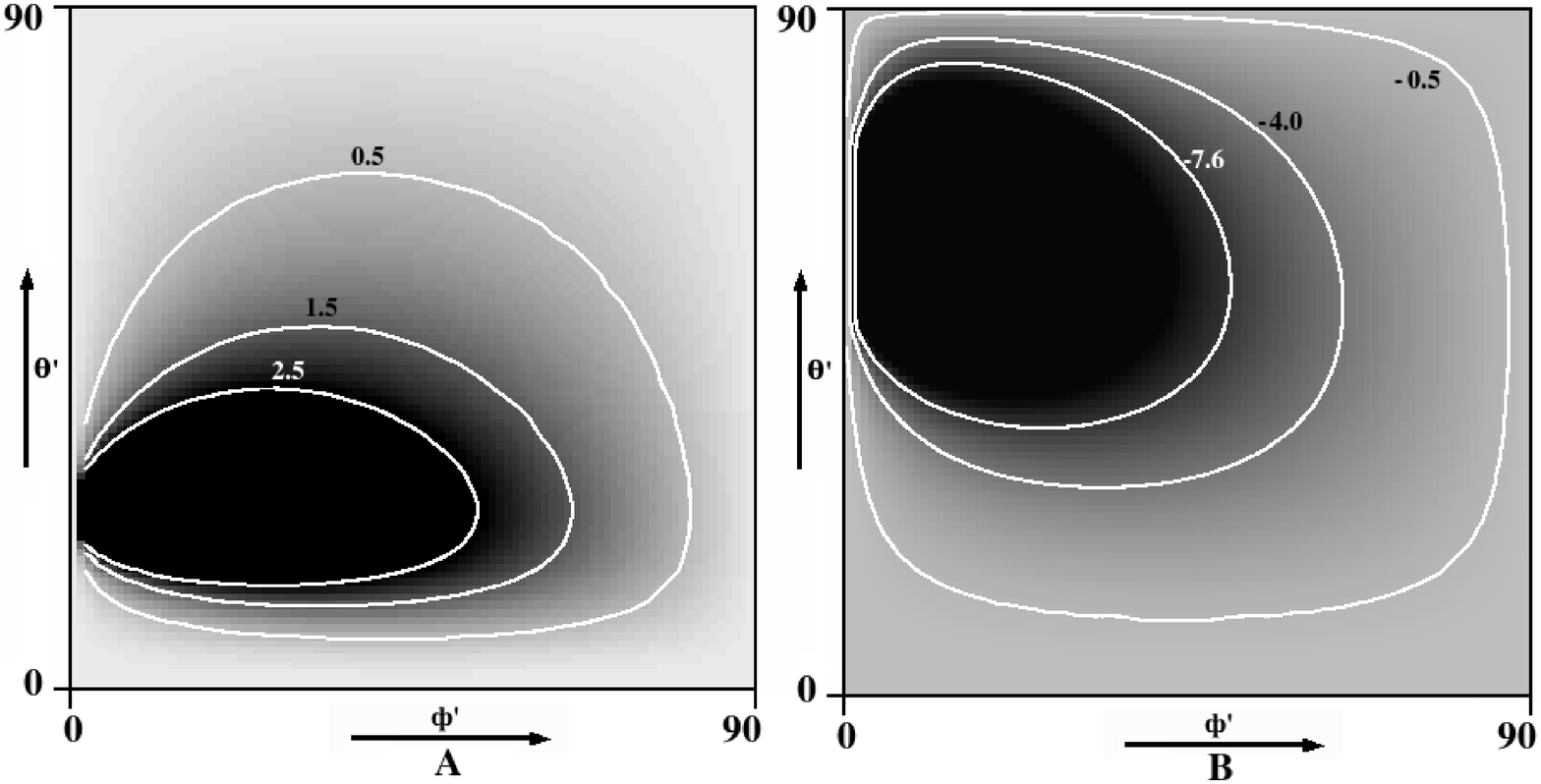}
 \caption{Contours of constant $\Theta_{d}$ for $fgh$ model with $\alpha=0$. Figure 1A is drawn with model parameters $(q_{0}, T_{0}, q_{\infty}, T_{\infty}) = (0.85, 0.2, 0.70, 0.3)$, while 1B is drawn with parameters $(0.85, 0.5, 0.70, 0.2)$.}
\end{figure}

To gain some further insight in the kind of results that we expect, we plot the constant $\Theta_{d}$ contours as functions of $(\theta{'}, \phi^{'})$, for models with chosen parameters. Figure 1, presents such constant $\Theta_{d}$ contours for $fgh$ models. Figure 1A is plotted with parameters $(q_{0}, T_{0}, q_{\infty}, T_{\infty}) = (0.85, 0.2, 0.70, 0.3)$, while figure 1B is drawn for model parameters $(q_{0}, T_{0}, q_{\infty}, T_{\infty}) = (0.85, 0.5, 0.70, 0.2)$. Eye inspection of these plots (and the similar plots for other choices of the model parameters) indicates the possible range of values of $(\theta^{'}, \phi^{'})$ where the observed $\Theta_{d}$ of a galaxy is close to the calculated $\Theta_{d}$ from a model. In this range, we shall obtain higher values of the likelihood. In particular, we find that the contours of small $\Theta_{d}$ are close to one of the principle planes.

\section{Method}
We apply the methodology outlined in section 1 to determine the orientations of NGC $7619$ and two synthetic galaxies. 
\subsection{NGC 7619}
The observed data of NGC $7619$ is $\epsilon_{in} = 0.28$ $\epsilon_{out} = 0.19$, $\theta_{d} = 4^{o}.0$ at $R_{in} =10.55$ arcsec and $R_{out} = 45.71$ arcsec. Effective radius $R_{e}$ of NGC 7619 is $32.0$ arcsec. This data is obtained from R-band photometry of Franx et al. (1989). While applying this data to calculate the probability of orientation, we consider the uncertainty of $0.02$ in ellipticities, and the uncertainty of $1^{o}.0$ in position angles both at $R_{in}$ and $R_{out}$.

Figure 2 presents the plot of the posterior probability distribution $\cal{P}$ of the orientation as a function of polar coordinates $(\theta^{'},\phi^{'})$ of the line of sight. We use a rectangular plot with $\theta^{'}$ going from $0^{o}$ to ${180^{o}}$ in the vertical direction from bottom to top and $\phi^{'}$ going from $0^{o}$ to $360^{o}$ in the horizontal direction from left to right. We label the eight octants of the space of the parameters $(\theta^{'}, \phi^{'})$, for some convenience in discussions. These labels are shown in figure 2. The origin of the coordinate system is 'placed' at the centre and the axes are 'drawn' along the principle axes of the triaxial galaxy. Here, we have used $fgh$ models with $\alpha = 0$ and have used $\epsilon_{in}$, $\epsilon_{out}$ and $\Theta_{d}$. The probability is shown in dark-grey shade : darker is the shade, higher is the probability. The white contour encloses 68$\%$ of the total probability which can be interpreted as 1$\sigma$ error bar.  
\begin{figure}
\centering
\includegraphics[width=9.0cm, angle=0]{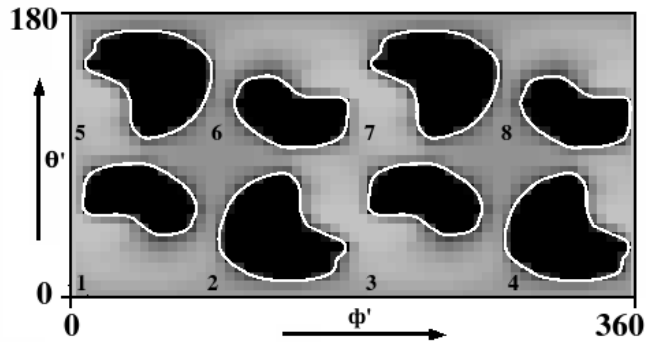}
 \caption{Plot of the posterior probability distribution $\cal{P}$ as a function of $(\theta^{'}, \phi^{'})$ for NGC 7619. Labels 1 to 8 denote the octants (see, text for detail).}
\end{figure}
\begin{figure}
\centering
\includegraphics[width=9.0cm, angle=0]{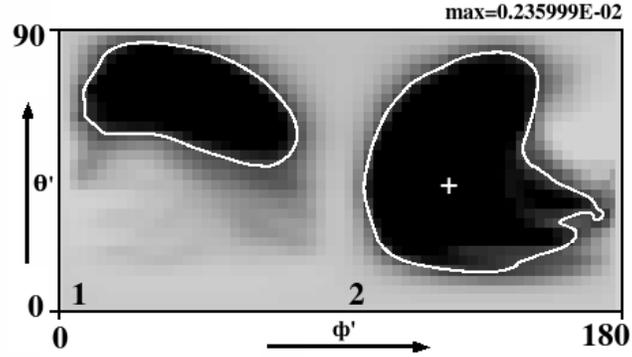}
 \caption{Plot of the posterior probability distribution $\cal{P}$ as a function of $(\theta^{'}, \phi^{'})$ for NGC 7619 in octants (1, 2). Plus marks the location of the maximum probability}
\end{figure}
\begin{figure}
\centering
\includegraphics[width=9.0cm, angle=0]{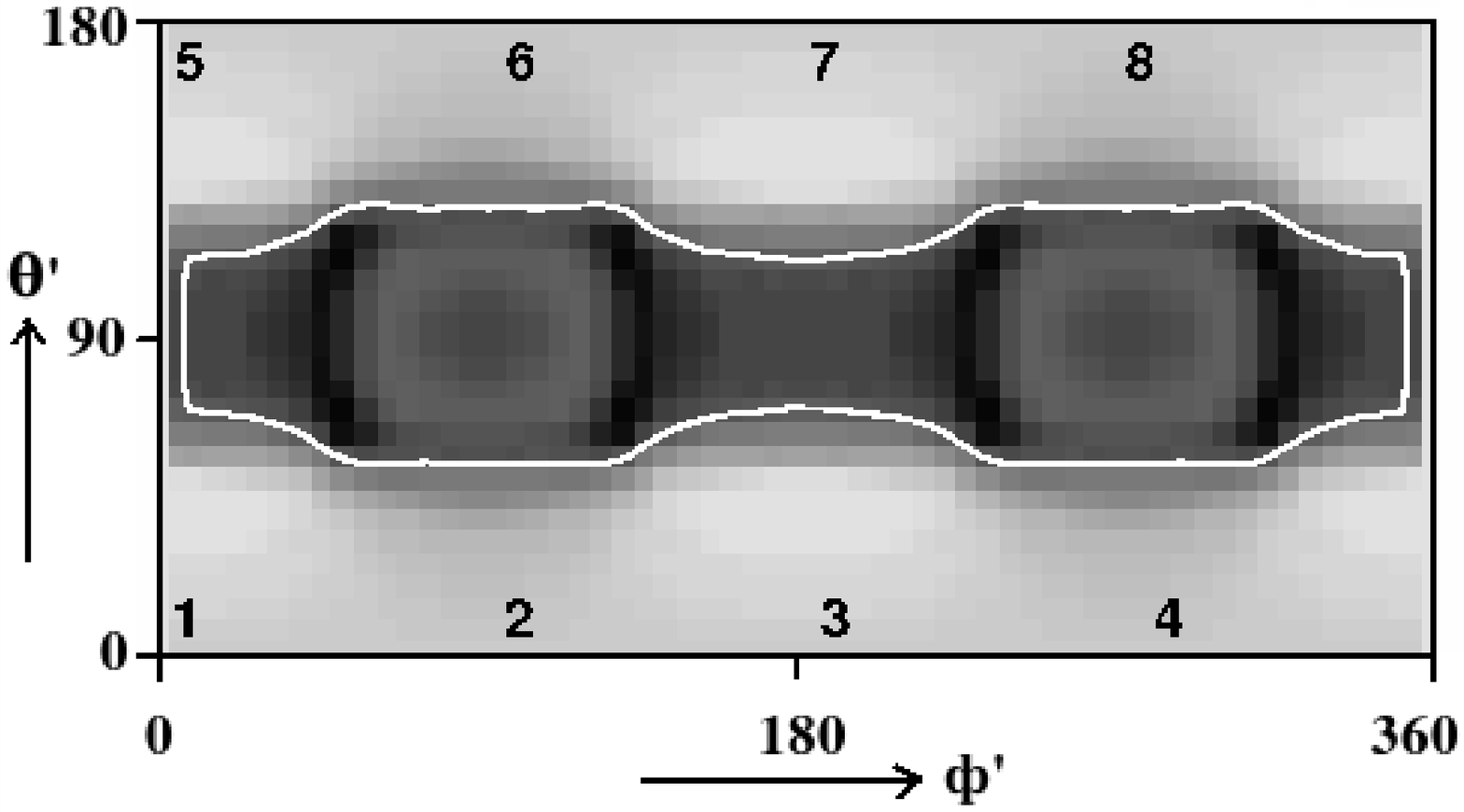}
 \caption{Plot of the posterior probability distribution $\cal{P}$ as a function of $(\theta^{'}, \phi^{'})$ for NGC 7619 using $\epsilon_{in}$ and $\epsilon_{out}$ only.}
\end{figure}

It is seen that the higher probability regions are disjoint blobs, each confined to one of the octants of the parameter space. Further, the plot is quite symmetric. Part of the plot in the pair of octants (1, 2) is repeated in the pairs (3, 4), (5, 6) and (7, 8). We thus conclude that the orientation is well constrained and it suffices to consider the probability distribution $\cal{P}$ in one of these pairs of octants only.

In figure 3, we present the plot of probability $\cal{P}$ in octants (1, 2). Here, we use $M^{2}$ models with $\beta = 0.2, 1.0,$ and $5.0$ and $fgh$ models with $\alpha = 0.0, 2.5$ and $5.0$ and take the unweighted sum of the probability over all these models. We find that the $1\sigma$ region occupies $27.9\%$ of the total parameter space. The probability is likelihood dominated and is well constrained. The location of the maximum probability is at $\theta^{'} = 40^{o}.5, \phi^{'} = 106^{o}.5$. In addition to above, the $1\sigma$ region extends to 3 more pairs of octants, as described above and exhibited through figure 2.

In obtaining the plots of figures 2 and 3, we have used $\epsilon_{in}, \epsilon_{out}$ and $\Theta_{d}$. We find that position angle difference $\Theta_{d}$ has a very important role. To illustrate this, we present the plot of $\cal{P}$ of NGC 7619 in figure 4, wherein we use $\epsilon_{in}$ and $\epsilon_{out}$ only. We find that ellipticities alone do not constrain the orientation. In particular, almost all values of $\phi^{'}$ are in the higher probability region. In the orientation estimates presented hereafter, we use the complete set of variables, namely $\epsilon_{in}, \epsilon_{out}$ and $\Theta_{d}$. The position angle difference $\Theta_{d}$ plays a major role in obtaining a likelihood dominated orientation estimates whereas the ellipticities $\epsilon_{in}$ and $\epsilon_{out}$ and the choices of $R_{in}$ and $R_{out}$ produce a fine tunning of the posterior probability distribution.  

\subsection{Synthetic galaxies}
In section 3.1, we presented the orientation estimates of NGC 7619. The reliability of the results can be tested by comparing them with results of the other workers. Alternatively, one can test of the results by recovering the orientation of an observed model from the ensemble. We consider the $M^{2} (\beta = 1, \gamma = 1.5)$ model with parameters $q_{0} = 0.9, T_{0} = 0.1, q_{\infty} = 0.5,  T_{\infty} = 0.2 $ and project it at viewing angles $\theta^{'} = 30^{o}.0$ and $\phi^{'} = 45^{o}.0$. We refer to this as galaxy A. The "observed" parameter are $\epsilon_{in} = 0.0408, \epsilon_{out} = 0.0899, \Theta_{in} = 11^{o}.6,$ and $\Theta_{out} = 17^{o}.4$ at $R_{in} = 0.25R_{e}$ and $R_{out} = 1.40R_{e} (R_{e} = 1.28)$. Here $\Theta_{d}$ is $5^{o}.8$. The uncertainties in ellipticities and in position angles are taken as the same as in the previous case of NGC 7619. Using $\epsilon_{in}, \epsilon_{out}$ and $ \Theta_{d}$ the plot of $\cal{P}$ over the entire range of $(\theta^{'}, \phi^{'})$. It is qualitatively similar to figure 2, and therefore, similar to the case of NGC 7619, we conclude that it is sufficient to consider the probability in one of the pairs of octants. 

\begin{figure}
\centering
\includegraphics[width=9.0cm, angle=0]{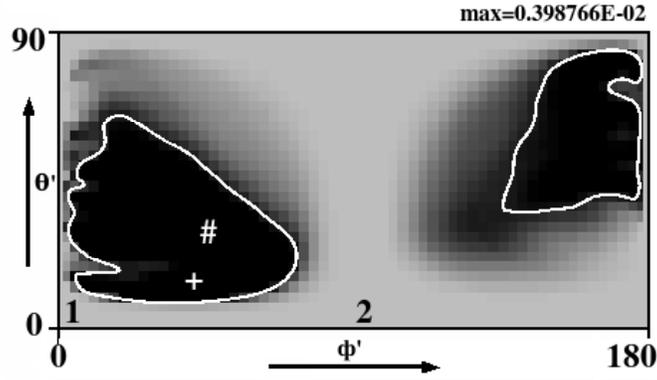}
 \caption{Same as Fig. 3, but for the synthetic galaxy A. The plus marks the location of highest $\cal{P}$, while the hedge marks the true orientation.}
\end{figure}
\begin{figure}
\centering
\includegraphics[width=9.0cm, angle=0]{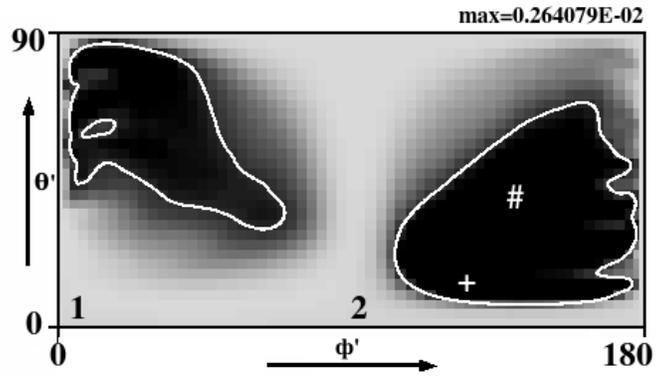}
 \caption{Same as Fig. 3, but for the synthetic galaxy B. The plus marks the location of highest $\cal{P}$, while the hedge marks the true orientation.}
\end{figure}

Figure 5 presents the plot of $\cal{P}$ to galaxy A in octants (1, 2). The contour encloses the region of $68\%$ of the total of probability. The hedge marks the true orientation and the plus marks the location of the highest probability. We find a good agreement between the results of probability calculations and the true orientation. This can be regarded as a proof of the reliability of the methodology.

As another example of synthetic galaxy, we consider $M^{2}$ model with the same values of $(q_{0}, T_{0}, q_{\infty}, T_{\infty})$ as those for galaxy A, but orientated at $\theta^{'} = 40^{o}.0$ and $\phi^{'} = 150^{o}.0$. The "observed" parameters are $\epsilon_{in} = 0.0573, \epsilon_{out} = 0.1117, \Theta_{in} = -6^{o}.4$ and $\Theta_{out} = -10^{o}.9$. $R_{in}$, $R_{out}$ and $R_{e}$ are the same as in galaxy A. The position angle difference $\Theta_{d}$ is $-4^{o}.5$. We refer to this as galaxy B. The plot of the probability $\cal{P}(\theta^{'}, \phi^{'})$ of galaxy B in octants 1 and 2, is presented in figure 6. Here again we find that the true orientation (marked by hedge) and the location of maximum probability (marked by plus) are close to each other.

The results of the galaxies A and B indicate that the orientation estimates of the galaxies using photometry alone are reliable. We find that galaxies A and B, as well as NGC 7619 exhibit large position angle difference ($|\Theta_{d}| \sim 4^{o}.0$ or more) and the orientations of these galaxies are well constrained.  

\section{Orientations of a sample of elliptical galaxies}
We now apply the methodology, which is now matured enough through its application to NGC 7619 and synthetic galaxies A and B, to a sample of elliptical galaxies. We find it convenient to divide the sample in categories based on the values of the position angle difference. This is based on large, moderate and small values of $|\Theta_{d}|$, without any sharp boundaries between them. NGC 7619 and the synthetic galaxies presented in section 3 are placed in large $|\Theta_{d}|$ category. Table 2 presents the list of elliptical galaxies of our sample, alongwith the observed data used in the models. Here L, M and S denote large, moderate and small $|\Theta_{d}|$ categories.

We begin by examining the orientations of the galaxies with large $|\Theta_{d}|$. Galaxies other than NGC 7619, having large $|\Theta_{d}|$ are NGC 1407, 2986, 4374, 4486 and 5638. We present the plots of $\cal{P}$ of NGC 1407, 2986, 4374, 4486 and 5638 in octants $(1, 2)$ in figures 7 - 11. We find that the orientations of galaxies are well constrained. We present a summary of the orientations of these galaxies in Table 3.
\begin{table*}
\centering
\caption{Observational data used in models.}
  \begin{tabular}{@{}ccccccccccc@{}}
  \hline
Galaxy & $R_{e}$ & $R_{in}$& $R_{out}$ & $ \epsilon_{in}$& $\epsilon_{out}$& $\Theta_{d}$  & category & Source\tablenotemark{a}\\
 \hline
NGC 1052 & 36.5 & 10.2  & 35.4  & 0.269 & 0.330 & 2.6  &  M & 1\\
NGC 1407 & 72.0 & 19.6  & 75.1  & 0.050 & 0.060 & -5.0 &  L & 2\\
NGC 2986 & 41.0 & 7.3   & 36.4  & 0.160 & 0.130 & 5.0  &  L & 2\\
NGC 3379 & 37.5 & 15.7  & 49.3  & 0.078 & 0.133 & 0.0  &  S & 1\\
NGC 4261 & 42.5 & 10.6  & 58.8  & 0.251 & 0.154 & 0.3  &  S & 1\\
NGC 4374 & 57.0 & 7.3   & 63.6  & 0.193 & 0.076 &-5.3  &  L & 1\\
NGC 4486 & 110.0& 15.0  & 69.0  & 0.033 & 0.100 &-13.3 &  L & 1\\
NGC 4551 & 21.0 & 5.3   & 32.4  & 0.227 & 0.284 & 1.0  &  S & 1\\
NGC 5638 & 29.5 & 8.0   & 36.9  & 0.056 & 0.118 & 8.4  &  L & 1\\
NGC 7619 & 32.0 & 10.5  & 45.7  & 0.280 & 0.190 & 4.0  &  L & 2\\
\hline
\end{tabular}
\tablenotetext{a}{(1) Peletier et al. (1990), (2) Franx et al. (1989).}
\end{table*}
\begin{table*}
\centering
  \caption{Summary of the orientation estimates.}
  \begin{tabular}{@{}cccc@{}}
  \hline
Galaxy & $\theta_{p}'$ & $\phi_{p}'$ & Percentage of the area of \\
       &       &     & $68\%$ highest probability \\ \hline
NGC 1052 & 79.5 & 139.5 & 34.1\%   \\
NGC 1407 & 13.5 & 127.5 & 33.0$\%$ \\
NGC 2986 & 31.5 & 118.5 & 27.4$\%$ \\
NGC 4374 & 67.5 & 175.5 & 26.5$\%$ \\
NGC 4486 & 73.5 & 10.5  & 11.0\%   \\
NGC 5638 & 16.5 & 37.5 & 17.0$\%$ \\
NGC 7619 & 40.5 & 106.5 & 29.7$\%$ \\
\hline
\end{tabular}
\end{table*}

\begin{figure}
\centering
\includegraphics[width=9.0cm, angle=0]{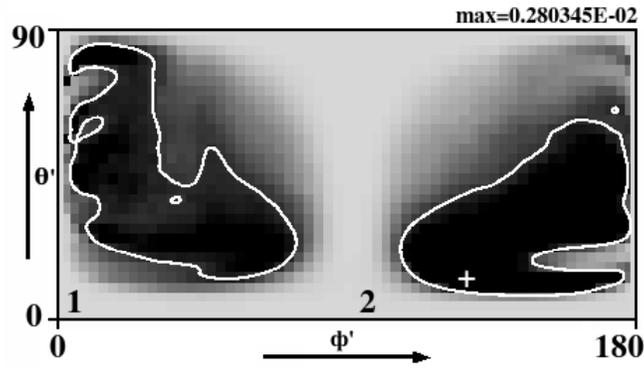}
 \caption{Same as Fig. 3, but for NGC 1407.}
\end{figure}
\begin{figure}
\centering
\includegraphics[width=9.0cm, angle=0]{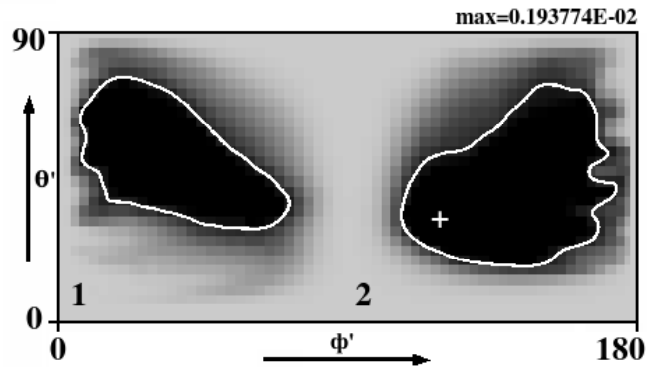}
 \caption{Same as Fig. 3, but for NGC 2986.}
\end{figure}
\begin{figure}
\centering
\includegraphics[width=9.0cm, angle=0]{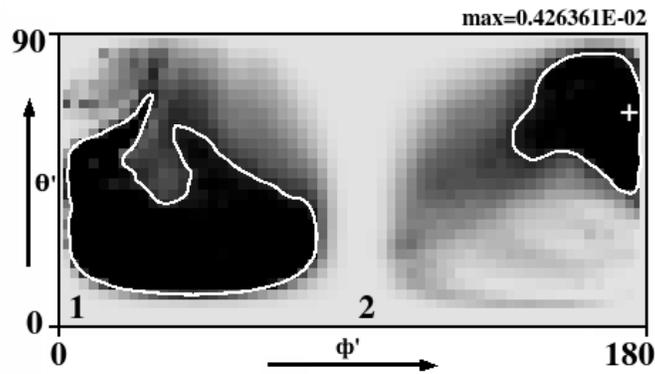}
 \caption{Same as Fig. 3, but for NGC 4374.}
\end{figure}
\begin{figure}
\centering
\includegraphics[width=9.0cm, angle=0]{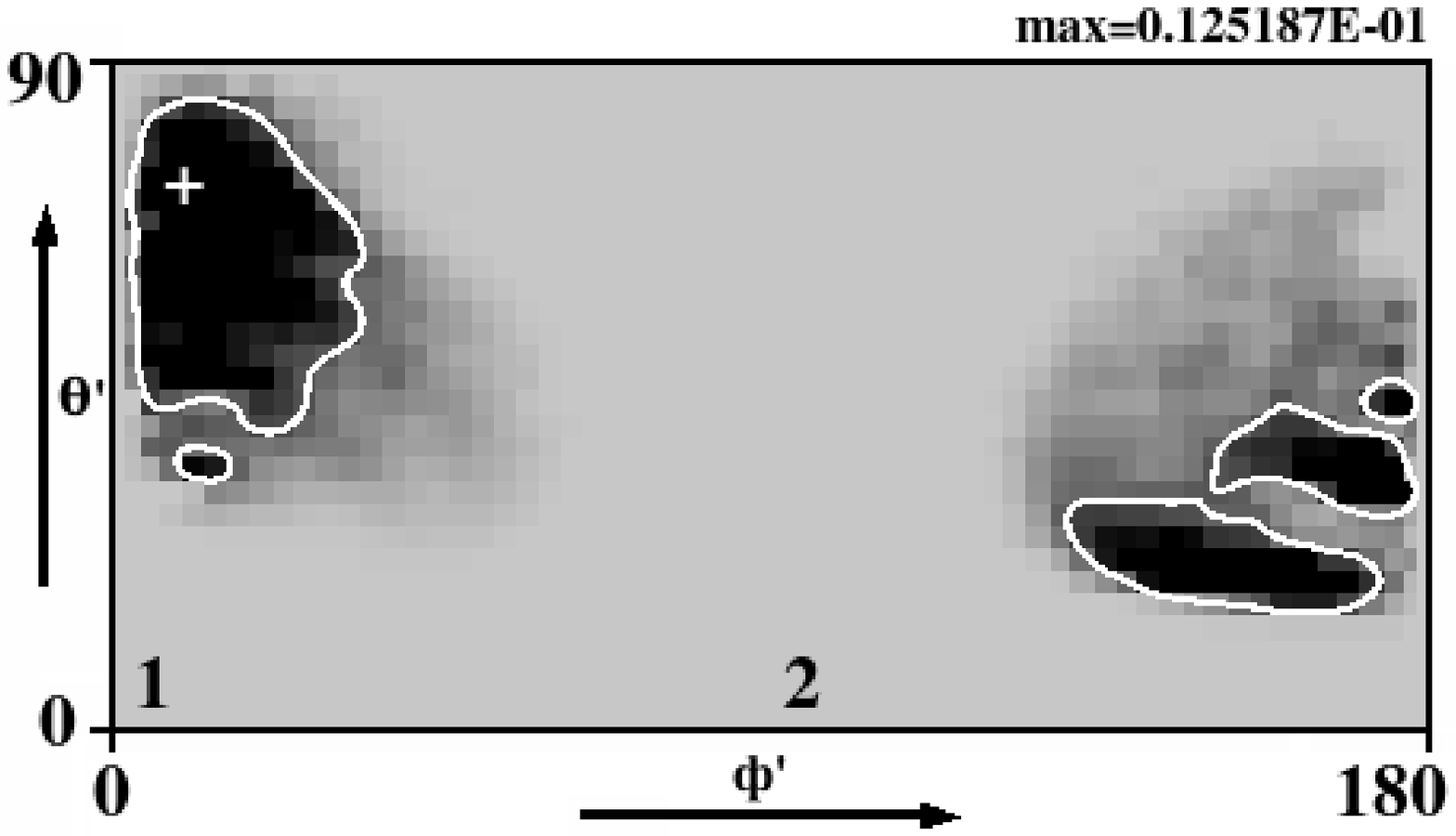}
 \caption{Same as Fig. 3, but for NGC 4486.}
\end{figure}
\begin{figure}
\centering
\includegraphics[width=9.0cm, angle=0]{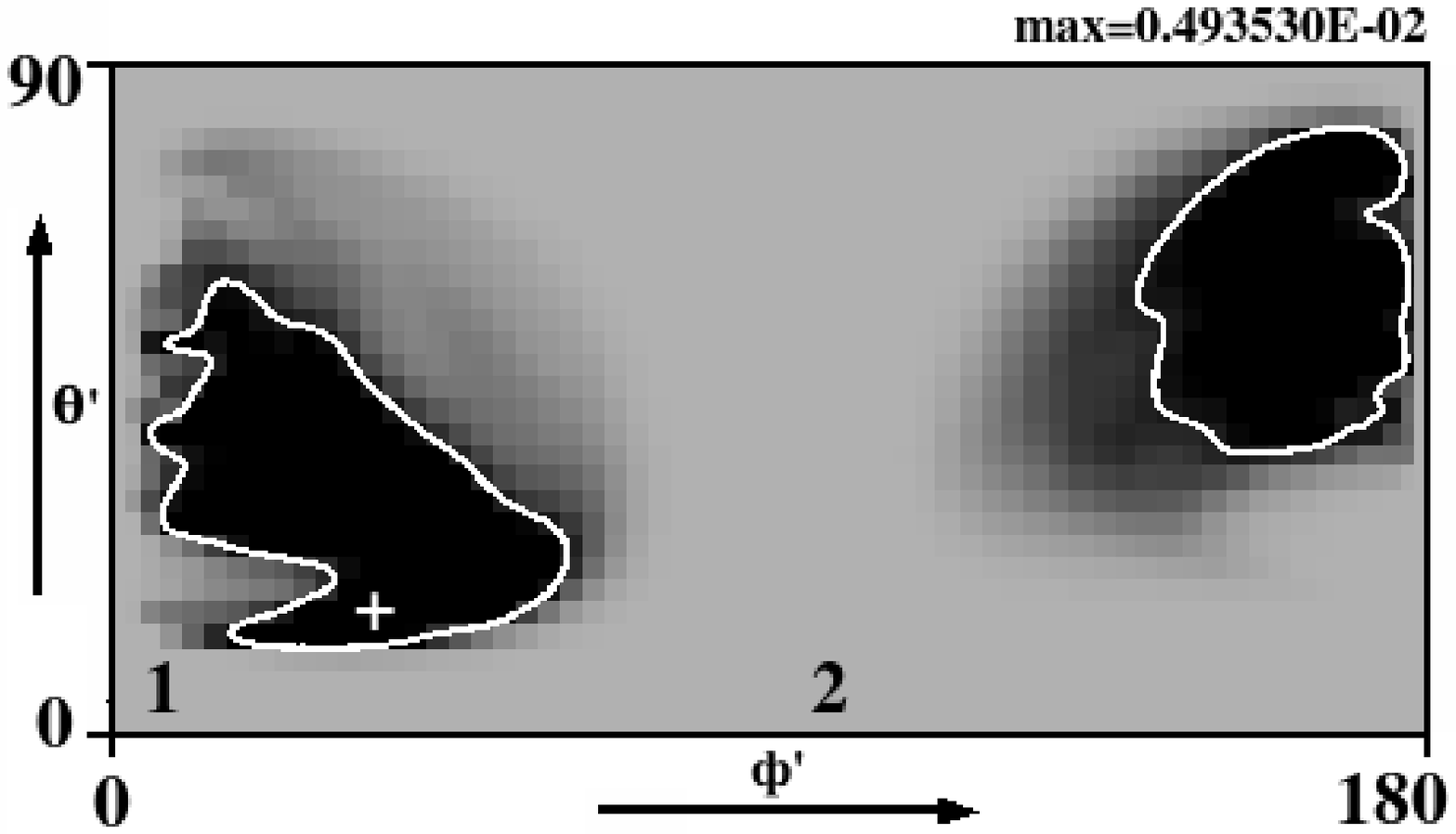}
 \caption{Same as Fig. 3, but for NGC 5638.}
\end{figure}
Next, we consider galaxies with small $|\Theta_{d}|$. Galaxies NGC 3379, 4261 and 4551 are placed in this category. Plot of the posterior density of NGC 3379 in the entire range of $(\theta^{'}, \phi^{'})$ is presented in figure 12. Although it is not constrained (almost all value of $\phi^{'}$ are in $1\sigma$ regions), but the symmetry of the plots in pairs of octants is clearly seen. Therefore, it is sufficient to consider the probability in octants 1 and 2 only.  Plots of the probability of the orientation of galaxies NGC 3379, 4261 and 4551 in octants (1, 2) are presented in figures 13, 14 and 15 respectively. We find that the orientation is not constrained, specifically, almost all values of $\phi^{'}$ are allowed in the high probability region. 

Galaxy NGC 1052 with $\Theta_{d} = 2^{o}.6$ is a border line case between the large and the small $|\Theta_{d}|$ categories. Probability density of the orientation of NGC 1052 is presented in figure 16. It is seen that the orientation is little poorly constrained : $68\%$ highest probability region extends over 2 octants of the viewing sphere. 
\begin{figure}
\centering
\includegraphics[width=9.0cm, angle=0]{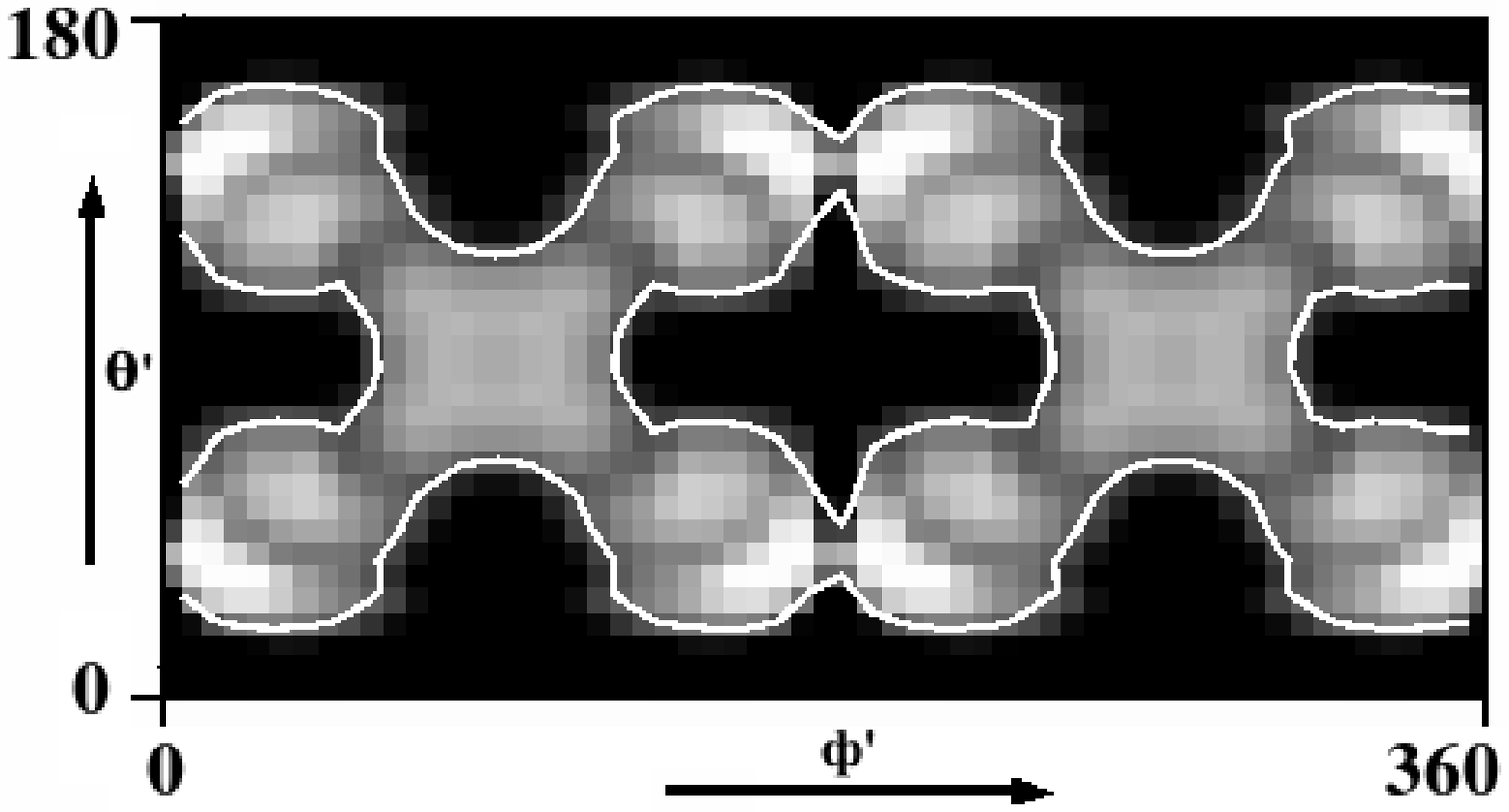}
 \caption{Same as Fig. 2, but for NGC 3379.}
\end{figure}
\begin{figure}
\centering
\includegraphics[width=9.0cm, angle=0]{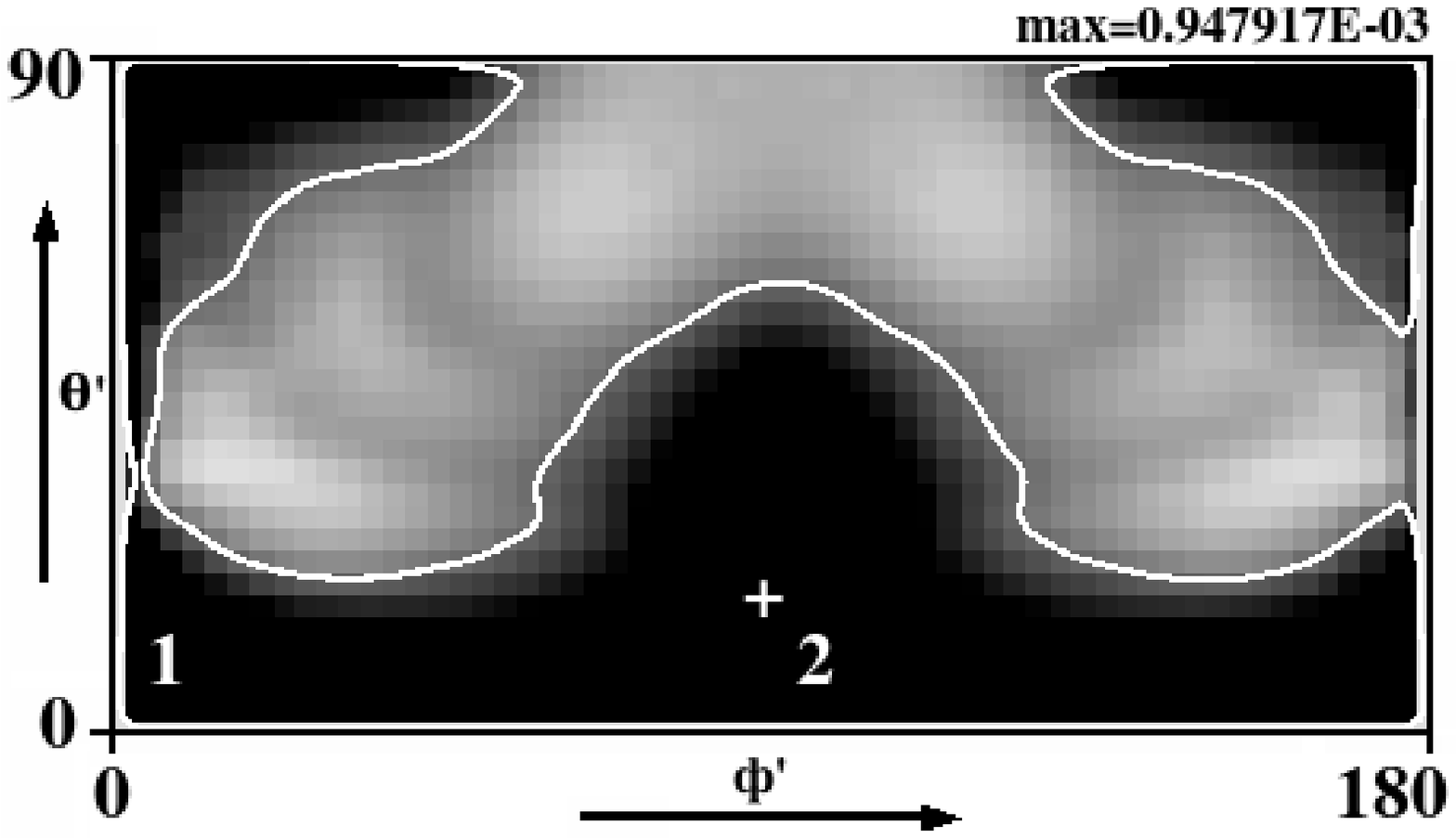}
 \caption{Same as Fig. 3, but for NGC 3379.}
\end{figure}
\begin{figure}
\centering
\includegraphics[width=9.0cm, angle=0]{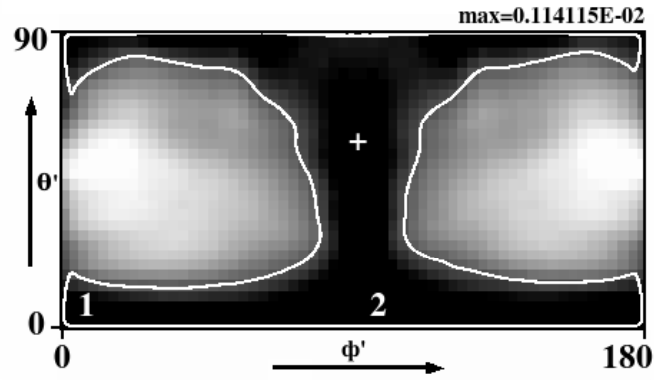}
 \caption{Same as Fig. 3, but for NGC 4261.}
\end{figure}
\begin{figure}
\centering
\includegraphics[width=9.0cm, angle=0]{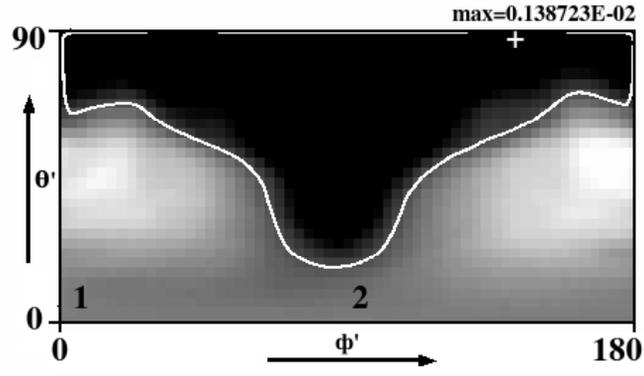}
 \caption{Same as Fig. 3, but for NGC 4551.}
\end{figure}
\begin{figure}
\centering
\includegraphics[width=9.0cm, angle=0]{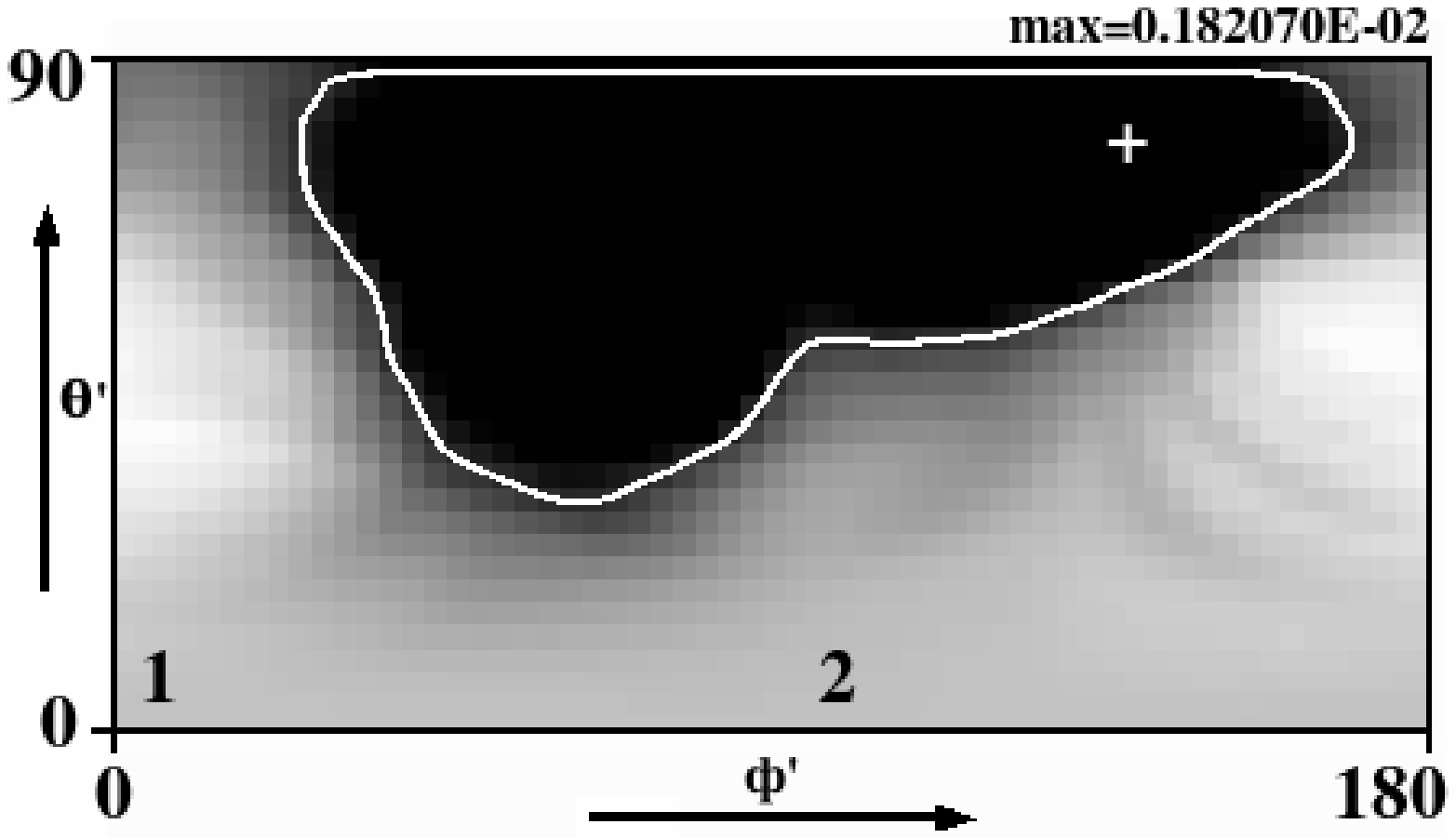}
 \caption{Same as Fig. 3, but for NGC 1052.}
\end{figure}

\section{Intrinsic alignment}
Intrinsic shapes of galaxies have implications for their formation and evolution, but the orientation of an arbitrarily chosen set of elliptical galaxies do not contain any useful physical information. Orientation of a very large number of ellipticals may be used to calculate an improved prior (Bak and Statler, 2000). In addition to above, we can also attempt to find correlations in orientations amongst the ellipticals of the same cluster. 

Presently, we discuss the alignment of the intrinsic principle axes of some ellipticals. It was \textit{assumed} in Statler et al. (1999) and in Statler (2001), while investigating the intrinsic shapes and orientations of NGC 1700 and 3379, that the intrinsic principle axes of these galaxies are aligned. We now wish to examine the consistency this assumption and our orientation estimates. 

We choose NGC 5638, 4486 and 3379. We consider each of these in different radial bins and estimate the orientations of these separately. Same orientation of different bins will imply the alignment (Appendix A, Statler et al. 1999). The radial bins and the respective observational data, used in orientation estimates, are shown in table 4. 
\begin{table*}
\centering
\caption{Observational data in various radial bins used in models.}
  \begin{tabular}{@{}ccccccc@{}}
  \hline
Galaxy & Bin & $R_{in}$ & $R_{out}$ & $\epsilon_{in}$ & $\epsilon_{out}$ &$\Theta_{d}$ \\ \hline
NGC 5638 & 1 & 8.0 & 36.9 & 0.056 & 0.118 & 8$^{o}$.4 \\
 	 & 2 & 4.7 & 36.9 & 0.048 & 0.118 & 10$^{o}$.4 \\
	 & 3 & 4.7 & 10.7 & 0.048 & 0.064 & 4$^{o}$.2 \\\hline
NGC 4486 & 1 & 15.0 & 69.0 &0.033 & 0.100 & -13$^{o}$.3 \\ 
	 & 2 & 15.0 & 122.2& 0.033 & 0.131 & -15$^{o}$.0 \\\hline
NGC 3379 & 1 & 15.7 & 49.3 & 0.078 & 0.133 & 0$^{o}$.0  \\
	 & 2 & 6.1  & 49.3 & 0.075 & 0.133 & -4$^{o}$.5  \\
	 & 3 & 6.1  & 15.7 & 0.075 & 0.078 & -4$^{o}$.7 \\\hline
\end{tabular}
\end{table*}

Note that the data in bin 1 of each of these galaxies are the same as used in section 4. Orientations of the bin 1 of NGC 4486, 5638 and 3379 are presented in figures 10, 11 and 13. The radial bin 2 of NGC 5638 engulfs the radial bin 1 and extends further at inner side. Orientation of bin 2 of NGC 5638 is presented in figure 17. We find that the plots of $\cal{P}$ of the orientation of bin 1 in figure 11 and of bin 2 in figure 17 are quite similar to each other. The peak of the orientation of bin 1 is at $(\theta^{'}_{p}, \phi^{'}_{p}) = (16^{o}.5, 37^{o}.5)$ which is close to the peak $(16^{o}.5, 40^{o}.5)$ of bin 2. These results support the notion that the intrinsic principle axes of NGC 5638 are aligned.
\begin{figure}
\centering
\includegraphics[width=9.0cm, angle=0]{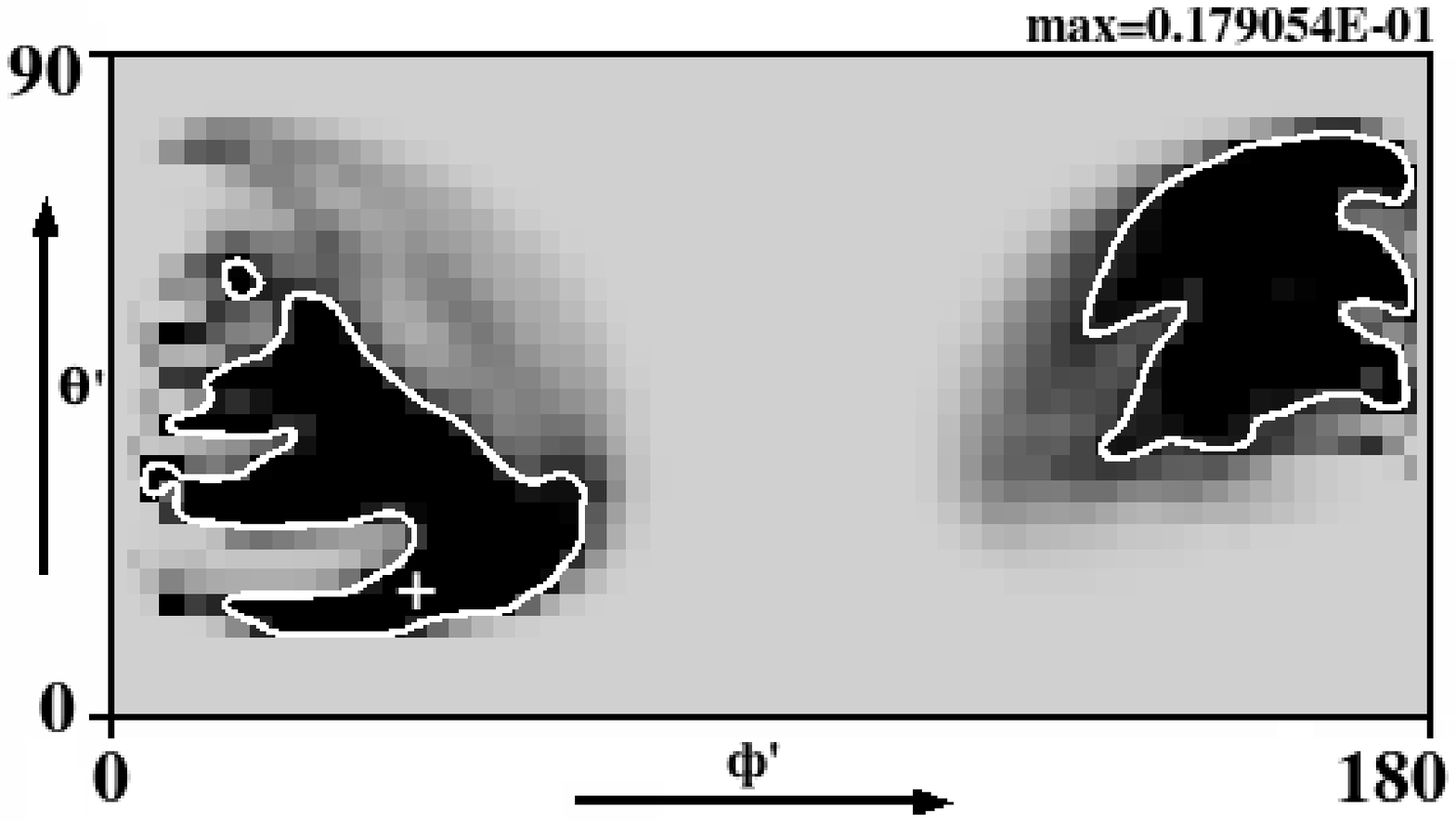}
 \caption{Same as fig. 3, but for bin 2 of NGC 5638.}
\end{figure}
\begin{figure}
\centering
\includegraphics[width=9.0cm, angle=0]{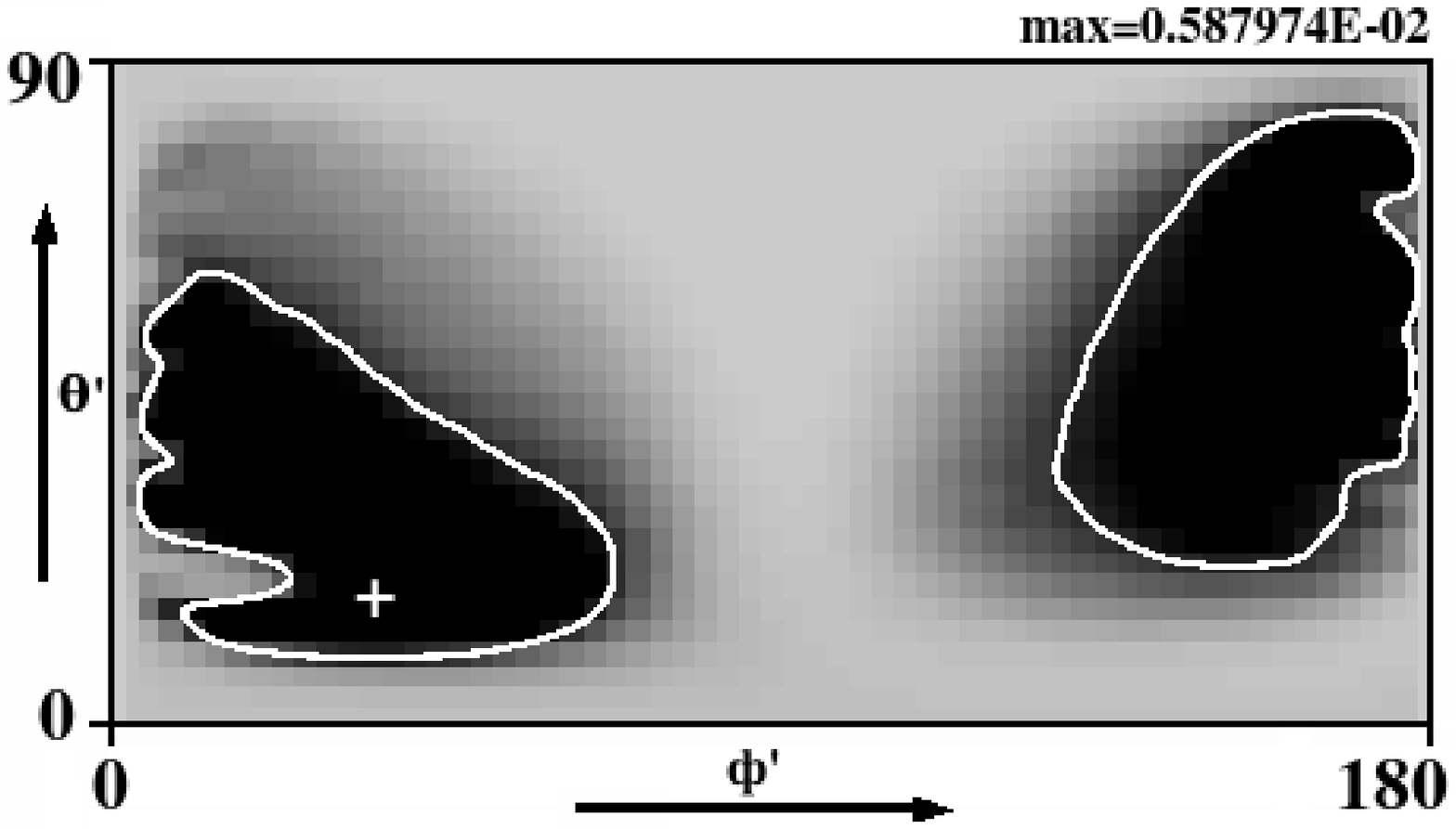}
 \caption{Same as fig. 3, but for bin 3 of NGC 5638.}
\end{figure}
\begin{figure}
\centering
\includegraphics[width=9.0cm, angle=0]{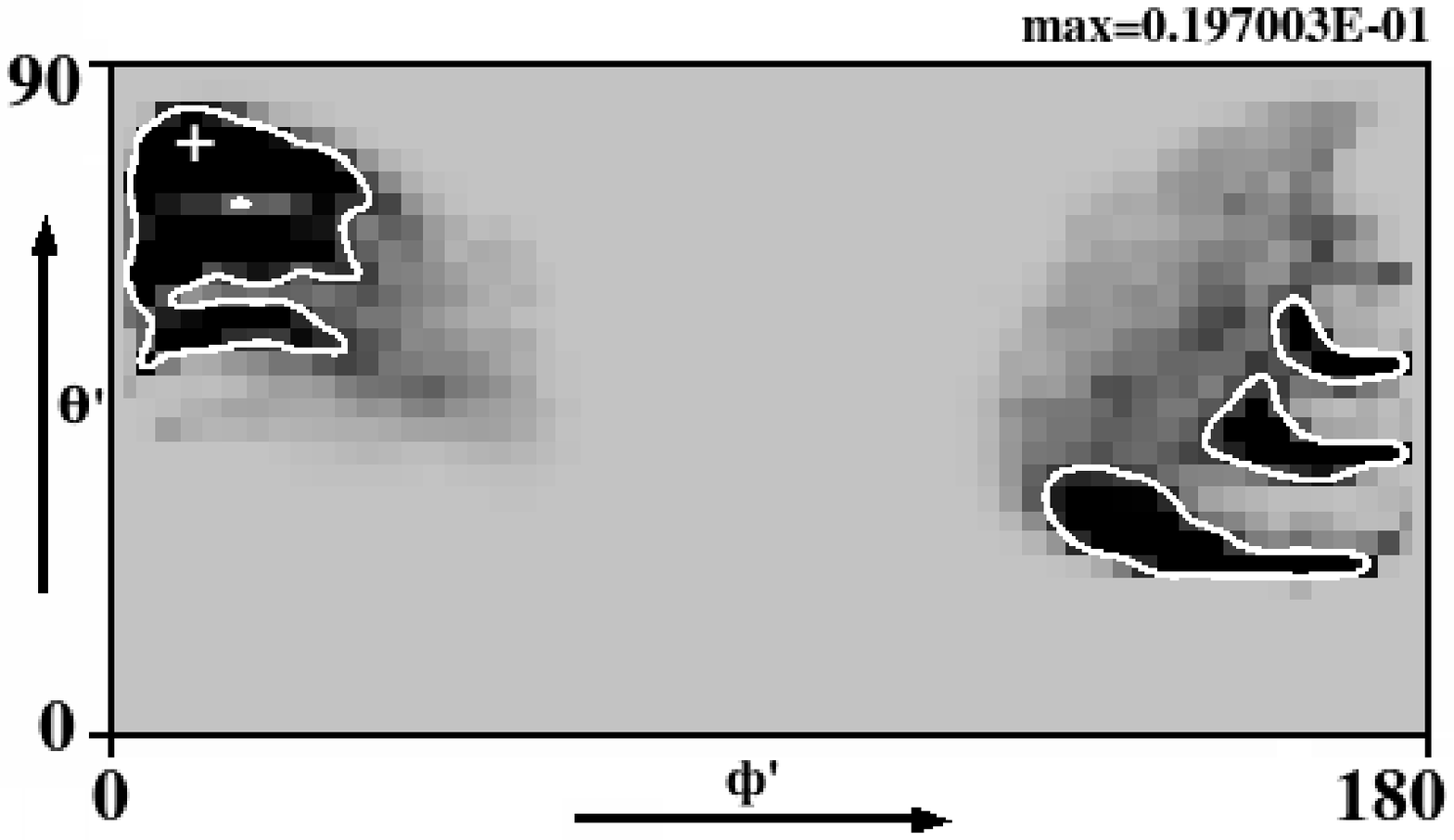}
 \caption{Same as fig. 3, but for bin 2 of NGC 4486.}
\end{figure}

To confirm the results further, we consider bin 3, which is a narrow and overlapping region between bin 1 and 2. The peak of the orientation of bin 3 is $(16^{o}.5, 37^{o}.5)$ which is the same as the peak orientation of bin 1. Plot of $\cal{P}(\theta^{'}, \phi^{'})$ of bin 3 is shown in figure 18, which is almost similar to plots of $\cal{P}$ of bin 1 and 2.
\begin{table*}
\centering
\caption{Summary of the orientations of various bins of NGC 5638, 4486 and 3379.}
\begin{tabular}{@{}ccccc@{}}
  \hline
Galaxy & Bin& $\theta_{p}'$ & $\phi_{p}'$ & Percentage of the area of \\
       & &       &     & $68\%$ highest probability \\ \hline
NGC 5638 & 1 & 16.5 & 37.5 & 17.0$\%$ \\
 	 & 2 & 16.5 & 40.5 & 13.9$\%$ \\
	 & 3 & 16.5 & 37.5 & 27.5\%\\\hline
NGC 4486 & 1 & 73.5 & 10.5 & 11.0$\%$ \\ 
	 & 2 & 73.5 & 10.5 & 8.1$\%$ \\\hline
NGC 3379 & 1 & 16.5 & 88.5 & 55.4$\%$ \\
	 & 2 & 19.5 & 109.5& 32.3\% \\
	 & 3 & 31.5 & 154.5& 23.7\% \\\hline
\end{tabular}
\end{table*}

A closer scruting of the plots of $\cal{P}$ reveal that while plots of bin 1 and bin 2 are very close to each other, plot of bin 3 is slightly different: $1\sigma$ area of the probability of bin 3 is $62\%$ higher than that of bin 1. We note that $R_{in}$ is very small and $R_{out}$ is very large for bin 1 and bin 2. However, $R_{out}$ of bin 3 is not very large. The values of $(\theta^{'}_{p}, \phi^{'}_{p})$ alongwith the area of $68\%$ highest probability of orientations of bins 1, 2, 3 are shown in table 5. Thus, the photometric data and the methodology are consistent with the idea that the intrinsic principle axes of NGC 5638 are aligned.

Probability  $\cal{P}$ of the orientation of bin 1 and bin 2 of NGC 4486 are presented in figures 10 and 19. The location of the maximum probability of bin 1 and bin 2 are almost the same. The results are consistent with the assumption of alignment of the intrinsic principle axes of NGC 4486. Note that both galaxies NGC 5638 and 4486, exhibit high values of $|\Theta_{d}|$ and their orientations are well constrained.
\begin{figure}[!ht]
\centering
\includegraphics[width=6.0cm, angle=0]{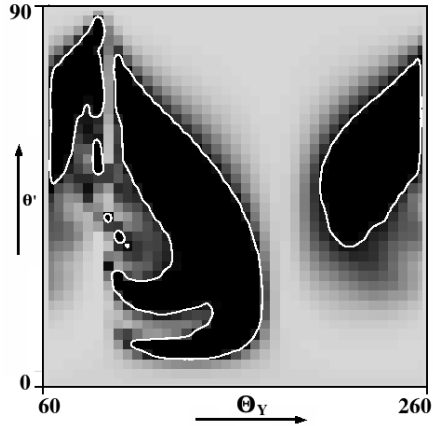}
 \caption{Plot of the probability distribution as a function of $(\theta',\Theta_{Y})$ for NGC 3379.}
\end{figure}

Results of NGC 3379 are complex. The results of the locations of maximum probability and of the area of $68\%$ highest probability of bins (1-3) are shown in table 5. We find that locations of maximum probability are not very close. Further, the orientation estimate of bin 1 is not very much likelihood dominated: $1\sigma$ and $2\sigma$ regions occupy $55.4\%$ and $92.1\%$, respectively.

Orientation of NGC 3379 is not well constrained even when the kinematical data is included (Statler, 2001, S94b). The probability distribution of NGC 3379 as a function of $\theta^{'}$ and the position angle $\Theta_{Y}$ of the projected y-axis of the body frame was presented in Statler (2001). Kinematical and photometric data were used and it was shown that while $\theta^{'}$ is not well constrained, $\Theta_{Y}$ is constrained. We re-examine this by using photometric models only, and using the data as quoted in Statler (2001). The plot is presented in figure 20, which can be compared with figure 5 (top) of Statler (2001). We find that the use of photometric data alone, worsens the situation\footnote{However, the photometry alone gives satisfactory shape estimates of NGC 3379 (C08)}. Here, both $\theta^{'}$ and $\Theta_{Y}$ are not well constrained.   

\section{Results and discussion}
We have demonstrated that the photometry can be used to obtain constraints on the orientations of the light distribution of elliptical galaxies, especially for the galaxies with large position angle difference. We have obtained the possible orientations of a few elliptical galaxies. We have also shown that the estimates are reliable by applying our methodology to the synthetic galaxies, wherein the orientations are known.

We have found that the posterior probability density of the orientation estimate is symmetric in pairs of octants (see, figure 2 and 12, and the relevant text). This should be  due to the mass models. The models are reflection symmetric about the principle planes.

We have shown that the intrinsic principle axes of NGC 5638 and 4486 are aligned by applying the orientation estimates using photometry. However, conclusions regarding the alignment of axes of NGC 3379 are not rigorously established. This is basically due to the reason that the orientation estimates of NGC 3379 are not likelihood dominated.  

The marginal posterior density of orientation obtained by integrating over the shape parameters gives the possible orientation which reproduces the photometric data most closely for the greatest variety of shapes. Likewise, the most probable shape obtained by marginalizing over the orientation reproduces the data over a greatest variety of orientations. These do not give simultaneous constraints of shape and orientation (Statler 2009, private communication).

\section*{Acknowledgments}
We are grateful to the anonymous reviewer for his critical comments and useful suggestions. This helped us to improve the paper in its present form.  We are greatly thankful to Prof. Thomas S. Statler for stimulating suggestions. DKC would like to thank the coordinator, IUCAA reference center, Pt. Ravishankar Shukla University Raipur, for the technical supports. AKD gratefully acknowledges the award of Rajiv Gandhi National Fellowship (SRF) of University Grant Commission.

\section*{References}
Bak J., \& Statler T.S. 2000, \aj, 120, 100 

 Binney J. 1985, \mnras, 212, 767 

 Chakraborty D.K. 2004 \aap, 423, 501 

 Chakraborty D.K., Singh A.K., \& Gaffar F. 2008, \mnras, 383, 1477 (C08) 

 Dehnen W. 1993, \mnras, 265, 250 

 de Zeeuw P.T., \& Carollo C.M. 1996, \mnras, 281, 1333 

 Franx M., Illingworth G.D., \& Heckman T. 1989, \aj, 98, 538 

 Peletier R.F., Davies R.L., Illingworth G.D., Davis L.E., \& Cawson M. 1990, \aj, 100, 1091 
 
 Stark A.A. 1977, \apj, 213, 368 

 Statler T.S. 1994a, \apj, 425, 500 (S94a) 

 Statler T.S. 1994b, \aj, 108, 111 (S94b) 

 Statler T.S. 2001, \aj, 121, 244 

 Statler T.S., Dejonghe H., \& Smecker-Hane T. 1999, \aj, 117, 126 

 Statler T.S., Emsellem E., Peletier R.F., \& Bacon R. 2004 \mnras, 253, 1 

 Tenjes P., Busarello G., Lango G., \& Zaggia S. 1993, \aap, 275, 61     

\label{lastpage}

\end{document}